\def\epsfigload{1}

\def\cerntp{1}

\ifnum\epsfigload=1
\documentstyle[epsfig,ifthen]{dgelsart}
\else
\documentstyle[ifthen]{dgelsart}
\newcommand{\psdraft}{}
\newcommand{\epsfig}[1]{}
\fi

\newcommand{\ifdraft}{nodraft}
\renewcommand{\ifdraft}{draft}

\newcommand{\ifpsdraft}{nodraft}

\newcommand{\vdate}{August 1995}
\newcommand{\cernnr}{95--232}
\newcommand{\beq}{\begin{equation}}
\newcommand{\eeq}{\end{equation}}
\newcommand{\beqn}{\begin{eqnarray}}
\newcommand{\eeqn}{\end{eqnarray}}

\newcommand{\porder}[1]{\mbox{${\cal O}(#1)$}}

\newcommand{\GeV}{\mbox{GeV}}

\newcommand{\pomeron}{Pomeron}

\newcommand{\fullline}{
\unitlength0.4mm
\begin{picture}(13,4)
\linethickness{0.3mm}
\put(-1,2.0){\line(1,0){15}}
\thinlines
\end{picture}
}

\newcommand{\dashline}{
\unitlength0.4mm
\begin{picture}(20,4)
\linethickness{0.3mm}
\put(-1,2.0){\line(1,0){4}}
\put(8,2.0){\line(1,0){4}}
\put(17,2.0){\line(1,0){4}}
\thinlines
\end{picture}
}

\newcommand{\dotline}{
\unitlength0.4mm
\begin{picture}(9,4)
\linethickness{0.3mm}
\put(-1,2.0){\line(1,0){1}}
\put(4,2.0){\line(1,0){1}}
\put(9,2.0){\line(1,0){1}}
\thinlines
\end{picture}
}

\newcommand{\longdashline}{
\unitlength0.4mm
\begin{picture}(22,4)
\linethickness{0.3mm}
\put(-1,2.0){\line(1,0){10}}
\put(13,2.0){\line(1,0){10}}
\thinlines
\end{picture}
}

\newcommand{\dashdotline}{
\unitlength0.4mm
\begin{picture}(17,4)
\linethickness{0.3mm}
\put(-1,2.0){\line(1,0){5}}
\put(8,2.0){\line(1,0){1}}
\put(13,2.0){\line(1,0){5}}
\thinlines
\end{picture}
}

\newcommand{\dotdotline}{
\unitlength0.4mm
\begin{picture}(12,4)
\linethickness{0.3mm}
\put(-1,2.0){\line(1,0){1}}
\put(2,2.0){\line(1,0){1}}
\put(9,2.0){\line(1,0){1}}
\put(12,2.0){\line(1,0){1}}
\thinlines
\end{picture}
}

\newcommand{\dashdotdotline}{
\unitlength0.4mm
\begin{picture}(11,4)
\linethickness{0.3mm}
\put(-1,2.0){\line(1,0){5}}
\put(8,2.0){\line(1,0){1}}
\put(11,2.0){\line(1,0){1}}
\thinlines
\end{picture}
}

\newcommand{\dashdashdotdotline}{
\unitlength0.4mm
\begin{picture}(22,4)
\linethickness{0.3mm}
\put(-1,2.0){\line(1,0){5}}
\put(9,2.0){\line(1,0){5}}
\put(19,2.0){\line(1,0){1}}
\put(22,2.0){\line(1,0){1}}
\thinlines
\end{picture}
}

\newcommand{\dashdotdotdotline}{
\unitlength0.4mm
\begin{picture}(15,4)
\linethickness{0.3mm}
\put(-1,2.0){\line(1,0){5}}
\put(9,2.0){\line(1,0){1}}
\put(12,2.0){\line(1,0){1}}
\put(15,2.0){\line(1,0){1}}
\thinlines
\end{picture}
}

\newcommand{\dgpicture}[2]{
\begin{picture}(#1,#2)
\thicklines
\thinlines
}

\newcommand{\chsign}[1]{%
{\ifthenelse{\equal{\ifdraft}{draft}}%
{
{\sf {\Large$\bullet$$\bullet$$\bullet$} #1 
                         {\Large$\bullet$$\bullet$$\bullet$} }%
}%
{}%
}}
\newcommand{\smallmark}[1]{
\marginpar{\fbox{\vspace{0.0cm}{\scriptsize #1}}}}

\newcommand{\labelm}[1]{%
\label{#1}%
\ifthenelse{\equal{\ifdraft}{draft}}%
{\smallmark{#1}}%
{}%
}

\newcommand{\labelmm}[1]{%
\label{#1}%
\ifthenelse{\equal{\ifdraft}{draft}}%
{\protect\fbox{\sf #1}}%
{}%
}

\newcommand{\beqm}[1]{%
\ifthenelse{\equal{\ifdraft}{draft}}%
{\smallmark{#1}}%
{}%
\beq \label{#1}}

\newcommand{\beqnm}[1]{%
\ifthenelse{\equal{\ifdraft}{draft}}%
{\smallmark{#1}}%
{}%
\beqn \label{#1}}

\ifthenelse{\equal{\ifpsdraft}{draft}}{\psdraft}{}

\catcode`\@=11

 \font\tenmsx=msam10 scaled \magstep1
 \font\sevenmsx=msam8
 \font\fivemsx=msam6
 \font\tenmsy=msbm10 scaled \magstep1
 \font\sevenmsy=msbm8
 \font\fivemsy=msbm6

\newfam\msxfam
\newfam\msyfam
\textfont\msxfam=\tenmsx  \scriptfont\msxfam=\sevenmsx
  \scriptscriptfont\msxfam=\fivemsx
\textfont\msyfam=\tenmsy  \scriptfont\msyfam=\sevenmsy
  \scriptscriptfont\msyfam=\fivemsy

\def\hexnumber@#1{\ifnum#1<10 \number#1\else
 \ifnum#1=10 A\else\ifnum#1=11 B\else\ifnum#1=12 C\else
 \ifnum#1=13 D\else\ifnum#1=14 E\else\ifnum#1=15 F\fi\fi\fi\fi\fi\fi\fi}

\def\msx@{\hexnumber@\msxfam}
\def\msy@{\hexnumber@\msyfam}
\mathchardef\boxdot="2\msx@00
\mathchardef\boxplus="2\msx@01
\mathchardef\boxtimes="2\msx@02
\mathchardef\square="0\msx@03
\mathchardef\blacksquare="0\msx@04
\mathchardef\centerdot="2\msx@05
\mathchardef\lozenge="0\msx@06
\mathchardef\blacklozenge="0\msx@07
\mathchardef\circlearrowright="3\msx@08
\mathchardef\circlearrowleft="3\msx@09
\mathchardef\rightleftharpoons="3\msx@0A
\mathchardef\leftrightharpoons="3\msx@0B
\mathchardef\boxminus="2\msx@0C
\mathchardef\Vdash="3\msx@0D
\mathchardef\Vvdash="3\msx@0E
\mathchardef\vDash="3\msx@0F
\mathchardef\twoheadrightarrow="3\msx@10
\mathchardef\twoheadleftarrow="3\msx@11
\mathchardef\leftleftarrows="3\msx@12
\mathchardef\rightrightarrows="3\msx@13
\mathchardef\upuparrows="3\msx@14
\mathchardef\downdownarrows="3\msx@15
\mathchardef\upharpoonright="3\msx@16

\mathchardef\downharpoonright="3\msx@17
\mathchardef\upharpoonleft="3\msx@18
\mathchardef\downharpoonleft="3\msx@19
\mathchardef\rightarrowtail="3\msx@1A
\mathchardef\leftarrowtail="3\msx@1B
\mathchardef\leftrightarrows="3\msx@1C
\mathchardef\rightleftarrows="3\msx@1D
\mathchardef\Lsh="3\msx@1E
\mathchardef\Rsh="3\msx@1F
\mathchardef\rightsquigarrow="3\msx@20
\mathchardef\leftrightsquigarrow="3\msx@21
\mathchardef\looparrowleft="3\msx@22
\mathchardef\looparrowright="3\msx@23
\mathchardef\circeq="3\msx@24
\mathchardef\succsim="3\msx@25
\mathchardef\gtrsim="3\msx@26
\mathchardef\gtrapprox="3\msx@27
\mathchardef\multimap="3\msx@28
\mathchardef\therefore="3\msx@29
\mathchardef\because="3\msx@2A
\mathchardef\doteqdot="3\msx@2B

\mathchardef\triangleq="3\msx@2C
\mathchardef\precsim="3\msx@2D
\mathchardef\lesssim="3\msx@2E
\mathchardef\lessapprox="3\msx@2F
\mathchardef\eqslantless="3\msx@30
\mathchardef\eqslantgtr="3\msx@31
\mathchardef\curlyeqprec="3\msx@32
\mathchardef\curlyeqsucc="3\msx@33
\mathchardef\preccurlyeq="3\msx@34
\mathchardef\leqq="3\msx@35
\mathchardef\leqslant="3\msx@36
\mathchardef\lessgtr="3\msx@37
\mathchardef\backprime="0\msx@38
\mathchardef\risingdotseq="3\msx@3A
\mathchardef\fallingdotseq="3\msx@3B
\mathchardef\succcurlyeq="3\msx@3C
\mathchardef\geqq="3\msx@3D
\mathchardef\geqslant="3\msx@3E
\mathchardef\gtrless="3\msx@3F
\mathchardef\sqsubset="3\msx@40
\mathchardef\sqsupset="3\msx@41
\mathchardef\vartriangleright="3\msx@42
\mathchardef\vartriangleleft="3\msx@43
\mathchardef\trianglerighteq="3\msx@44
\mathchardef\trianglelefteq="3\msx@45
\mathchardef\bigstar="0\msx@46
\mathchardef\between="3\msx@47
\mathchardef\blacktriangledown="0\msx@48
\mathchardef\blacktriangleright="3\msx@49
\mathchardef\blacktriangleleft="3\msx@4A
\mathchardef\vartriangle="3\msx@4D
\mathchardef\blacktriangle="0\msx@4E
\mathchardef\triangledown="0\msx@4F
\mathchardef\eqcirc="3\msx@50
\mathchardef\lesseqgtr="3\msx@51
\mathchardef\gtreqless="3\msx@52
\mathchardef\lesseqqgtr="3\msx@53
\mathchardef\gtreqqless="3\msx@54
\mathchardef\Rrightarrow="3\msx@56
\mathchardef\Lleftarrow="3\msx@57
\mathchardef\veebar="2\msx@59
\mathchardef\barwedge="2\msx@5A
\mathchardef\doublebarwedge="2\msx@5B
\mathchardef\angle="0\msx@5C
\mathchardef\measuredangle="0\msx@5D
\mathchardef\sphericalangle="0\msx@5E
\mathchardef\varpropto="3\msx@5F
\mathchardef\smallsmile="3\msx@60
\mathchardef\smallfrown="3\msx@61
\mathchardef\Subset="3\msx@62
\mathchardef\Supset="3\msx@63
\mathchardef\Cup="2\msx@64

\mathchardef\Cap="2\msx@65

\mathchardef\curlywedge="2\msx@66
\mathchardef\curlyvee="2\msx@67
\mathchardef\leftthreetimes="2\msx@68
\mathchardef\rightthreetimes="2\msx@69
\mathchardef\subseteqq="3\msx@6A
\mathchardef\supseteqq="3\msx@6B
\mathchardef\bumpeq="3\msx@6C
\mathchardef\Bumpeq="3\msx@6D
\mathchardef\lll="3\msx@6E

\mathchardef\ggg="3\msx@6F

\mathchardef\circledS="0\msx@73
\mathchardef\pitchfork="3\msx@74
\mathchardef\dotplus="2\msx@75
\mathchardef\backsim="3\msx@76
\mathchardef\backsimeq="3\msx@77
\mathchardef\complement="0\msx@7B
\mathchardef\intercal="2\msx@7C
\mathchardef\circledcirc="2\msx@7D
\mathchardef\circledast="2\msx@7E
\mathchardef\circleddash="2\msx@7F
\def\ulcorner{\delimiter"4\msx@70\msx@70 }
\def\urcorner{\delimiter"5\msx@71\msx@71 }
\def\llcorner{\delimiter"4\msx@78\msx@78 }
\def\lrcorner{\delimiter"5\msx@79\msx@79 }
\def\yen{\mathhexbox\msx@55 }
\def\checkmark{\mathhexbox\msx@58 }
\def\circledR{\mathhexbox\msx@72 }
\def\maltese{\mathhexbox\msx@7A }
\mathchardef\lvertneqq="3\msy@00
\mathchardef\gvertneqq="3\msy@01
\mathchardef\nleq="3\msy@02
\mathchardef\ngeq="3\msy@03
\mathchardef\nless="3\msy@04
\mathchardef\ngtr="3\msy@05
\mathchardef\nprec="3\msy@06
\mathchardef\nsucc="3\msy@07
\mathchardef\lneqq="3\msy@08
\mathchardef\gneqq="3\msy@09
\mathchardef\nleqslant="3\msy@0A
\mathchardef\ngeqslant="3\msy@0B
\mathchardef\lneq="3\msy@0C
\mathchardef\gneq="3\msy@0D
\mathchardef\npreceq="3\msy@0E
\mathchardef\nsucceq="3\msy@0F
\mathchardef\precnsim="3\msy@10
\mathchardef\succnsim="3\msy@11
\mathchardef\lnsim="3\msy@12
\mathchardef\gnsim="3\msy@13
\mathchardef\nleqq="3\msy@14
\mathchardef\ngeqq="3\msy@15
\mathchardef\precneqq="3\msy@16
\mathchardef\succneqq="3\msy@17
\mathchardef\precnapprox="3\msy@18
\mathchardef\succnapprox="3\msy@19
\mathchardef\lnapprox="3\msy@1A
\mathchardef\gnapprox="3\msy@1B
\mathchardef\nsim="3\msy@1C
\mathchardef\napprox="3\msy@1D
\mathchardef\varsubsetneq="3\msy@20
\mathchardef\varsupsetneq="3\msy@21
\mathchardef\nsubseteqq="3\msy@22
\mathchardef\nsupseteqq="3\msy@23
\mathchardef\subsetneqq="3\msy@24
\mathchardef\supsetneqq="3\msy@25
\mathchardef\varsubsetneqq="3\msy@26
\mathchardef\varsupsetneqq="3\msy@27
\mathchardef\subsetneq="3\msy@28
\mathchardef\supsetneq="3\msy@29
\mathchardef\nsubseteq="3\msy@2A
\mathchardef\nsupseteq="3\msy@2B
\mathchardef\nparallel="3\msy@2C
\mathchardef\nmid="3\msy@2D
\mathchardef\nshortmid="3\msy@2E
\mathchardef\nshortparallel="3\msy@2F
\mathchardef\nvdash="3\msy@30
\mathchardef\nVdash="3\msy@31
\mathchardef\nvDash="3\msy@32
\mathchardef\nVDash="3\msy@33
\mathchardef\ntrianglerighteq="3\msy@34
\mathchardef\ntrianglelefteq="3\msy@35
\mathchardef\ntriangleleft="3\msy@36
\mathchardef\ntriangleright="3\msy@37
\mathchardef\nleftarrow="3\msy@38
\mathchardef\nrightarrow="3\msy@39
\mathchardef\nLeftarrow="3\msy@3A
\mathchardef\nRightarrow="3\msy@3B
\mathchardef\nLeftrightarrow="3\msy@3C
\mathchardef\nleftrightarrow="3\msy@3D
\mathchardef\divideontimes="2\msy@3E
\mathchardef\varnothing="0\msy@3F
\mathchardef\nexists="0\msy@40
\mathchardef\mho="0\msy@66
\mathchardef\thorn="0\msy@67
\mathchardef\beth="0\msy@69
\mathchardef\gimel="0\msy@6A
\mathchardef\daleth="0\msy@6B
\mathchardef\lessdot="3\msy@6C
\mathchardef\gtrdot="3\msy@6D
\mathchardef\ltimes="2\msy@6E
\mathchardef\rtimes="2\msy@6F
\mathchardef\shortmid="3\msy@70
\mathchardef\shortparallel="3\msy@71
\mathchardef\smallsetminus="2\msy@72
\mathchardef\thicksim="3\msy@73
\mathchardef\thickapprox="3\msy@74
\mathchardef\approxeq="3\msy@75
\mathchardef\succapprox="3\msy@76
\mathchardef\precapprox="3\msy@77
\mathchardef\curvearrowleft="3\msy@78
\mathchardef\curvearrowright="3\msy@79
\mathchardef\digamma="0\msy@7A
\mathchardef\varkappa="0\msy@7B
\mathchardef\hslash="0\msy@7D
\mathchardef\hbar="0\msy@7E
\mathchardef\backepsilon="3\msy@7F
\def\Bbb{\ifmmode\let\next\Bbb@\else
 \def\next{\errmessage{Use \string\Bbb\space only in math mode}}\fi\next}
\def\Bbb@#1{{\Bbb@@{#1}}}
\def\Bbb@@#1{\fam\msyfam#1}

\catcode`\@=12
\font\teneusmf=eufm10 scaled 1200
\font\seveneusmf=eufm8
\font\fiveeusmf=eufm6
\newfam\eusmffam
\textfont\eusmffam=\teneusmf
\scriptfont\eusmffam=\seveneusmf
\scriptscriptfont\eusmffam=\fiveeusmf

\font\teneusm=eusm10 scaled 1200
\font\seveneusm=eusm8
\font\fiveeusm=eusm6
\newfam\eusmfam
\textfont\eusmfam=\teneusm
\scriptfont\eusmfam=\seveneusm
\scriptscriptfont\eusmfam=\fiveeusm

\font\teneusmc=cmsy10 scaled 1200
\font\seveneusmc=cmsy8
\font\fiveeusmc=cmsy6
\newfam\eusmcfam
\textfont\eusmcfam=\teneusmc
\scriptfont\eusmcfam=\seveneusmc
\scriptscriptfont\eusmcfam=\fiveeusmc

\newcommand{\markit}[1]{{\sc #1}}
\newcommand{\markh}[1]{\markboth{\markit{#1}}{\markit{#1}}}

\newcommand{\titletextCERN}
{
Estimating Diffractive Higgs Boson\\ \vspace{2mm}Production
at LHC from HERA Data
}

\newcommand{\titletextPL}
{
Estimating Diffractive Higgs Boson Production
at LHC\\ from HERA Data
}

\newcommand{\abstracttext}
{
Using a recently proposed factorization hypothesis for semi-inclusive hard
processes in QCD, one can study, in principle, the diffractive production
of the Standard Model Higgs boson at LHC using only, as input, $ep$
diffractive hard-processes data of the type recently collected and analyzed by
the H1 and ZEUS collaborations at HERA. While waiting for a more precise and
complete set of data, we combine here the existing data with a simple
\pomeron-exchange picture and find a large spread in the Higgs boson production
cross section, depending on the input parametrization of the \pomeron's parton
content. In particular, if the \pomeron{} gluon density $f_{g/{\Bbb P}}(\beta)$
is peaked at large $\beta$ for small scales, single diffractive events will
represent a sizeable fraction of all produced Higgs bosons with an expected
better-than-average signal-to-background ratio.
}

\begin{document}


\ifnum\cerntp=1

\thispagestyle{empty}

\renewcommand{\thefootnote}{\fnsymbol{footnote}}
\setcounter{footnote}{0}

\begin{flushright}
CERN--TH/\cernnr
\end{flushright}
 
\vspace{1.5cm}

\begin{center}

{\Large\bf \titletextCERN}



\vspace{0.8cm}

{\large\bf D.~Graudenz}\footnote[1]{\em Electronic
mail address: Dirk.Graudenz\char64{}cern.ch}\footnote[2]{\em
WWW URL: http://surya11.cern.ch/users/graudenz/} 
{\large\bf and G.~Veneziano}\footnote[3]{\em Electronic
mail address: venezia\char64{}nxth04.cern.ch}

\vspace{0.5cm}

{\large\it Theoretical Physics Division, CERN\\
\vspace{0.0cm}
CH--1211 Geneva 23\\
\vspace{0.1cm}
Switzerland}

\end{center}

\vspace{1.0cm}

\begin{center}
{\bf Abstract}
\end{center}

\hspace{4mm}
\abstracttext

\vfill
\noindent
CERN--TH/\cernnr\\
\vdate

\clearpage
\setcounter{page}{1}

\renewcommand{\thefootnote}{\arabic{footnote}}
\setcounter{footnote}{0}

\fi
 

\ifnum\cerntp=0

\begin{frontmatter}

\title{\titletextPL}

\author[CERN]{D.~Graudenz\thanksref{EMD1}\thanksref{EMD2}} and
\author[CERN]{G.~Veneziano\thanksref{EMV}}

\address[CERN]{Theoretical Physics Division, CERN,
CH--1211 Geneva 23, Switzerland}

\thanks[EMD1]{\em Electronic
mail address: Dirk.Graudenz\char64{}cern.ch
}

\thanks[EMD2]{
\em WWW URL: http://surya11.cern.ch/users/graudenz/}

\thanks[EMV]{\em Electronic
mail address:
venezia\char64{}nxth04.cern.ch}

\begin{abstract}
\abstracttext
\end{abstract}

\end{frontmatter}

\fi

Searching for the Higgs boson is one of the main goals of future
hadron colliders.
The main search strategy for this elusive particle
rests on the production via 
the dominant $gg\rightarrow H$ fusion process, and on the observation of the
subsequent decays into $ZZ$ and $ZZ^*$, for the mass ranges
$m_H>2m_Z$ and $130\,\GeV<m_H<2m_Z$, respectively
(the experimental signature being four charged leptons from the 
$Z$, $Z^*$ decays), or of
the rare decay into two photons, $\mbox{H}\rightarrow\gamma\gamma$,
for the intermediate mass range $90\,\GeV<m_H<130\,\GeV$. 
The search for the Higgs boson in the intermediate mass range is believed to be
very difficult \cite{1}.
The process $pp\rightarrow gg+X\rightarrow H+X\rightarrow\gamma\gamma+X$
suffers indeed from the large irreducible background
$pp\rightarrow\gamma\gamma+X$
and 
from the reducible background $pp\rightarrow
\mbox{jet}+\mbox{jet}+X$, $\mbox{jet}+\gamma+X$, 
where the jets fake photons
\cite{2,3}.

Additional information on the parton-level initial state 
(i.e. distinguishing \ $gg$, $qg$ or $q\overline{q}$ induced processes) 
in a selected class of
Higgs boson production events may help to reduce the background. For instance, 
the irreducible two-photon background
is mainly induced by a quark--antiquark pair and is therefore suppressed
if  one is able to bias the sample so as to favour initial state gluons 
(see the brief discussion of backgrounds at the end of the paper).
In hard diffractive processes the proton stays essentially intact and its
valence quarks go straight into the leading final proton.
What recoils against the proton--proton system is naturally expected 
to be gluon-rich, at least in comparison to unbiased 
(i.e. diffractive plus non-diffractive) events.

Diffractive hard processes are usually described in the framework of
QCD by introducing  an effective
flux for the exchanged object with vacuum quantum numbers
(for short, called the ``\pomeron'' (${\Bbb P}$) in the sequel)  
and by parametrizing the parton densities of the \pomeron{} itself
\cite{4}. There is some experimental evidence in support of
the theoretical expectation that
the \pomeron{} is a gluon-rich object \cite{5,6,7}. 
Tagging leading protons
with a large momentum \cite{8}, or observing gaps in rapidity
as an experimental sign for diffractive processes,
may allow to exploit
additional information about the hard scattering process
and to enrich the sample of gluon-initiated events.

In this paper we study the production of the Standard Model 
Higgs boson in 
hard diffractive processes at LHC, 
using experimental fits to the
\pomeron{} parton densities from groups working at HERA.
We point out, however, that,
for a study of hard 
diffractive events, this (partly model dependent) 
theoretical framework is not really needed. A direct 
parametrization of the gluon content of a proton fragmenting into a
leading proton would be sufficient and is desirable in order
to facilitate forthcoming studies of hard diffraction. 
The corresponding theoretical framework
of {\em fracture functions}
has been developed recently \cite{9,10}. 
The fracture function $M^i_{p,h}(x,z,Q^2)$ gives the
joint probability distribution for an observed hadron $h$
(e.g.\ a leading proton) with a specific momentum fraction $z$ in the proton
fragmentation region, and a parton $i$ (e.g.\ a gluon) 
of momentum fraction $x=\beta(1-z)$ initiating the
hard process. The functions $F_2^{D(3)}(\beta,Q^2,x_{\Bbb P})$
introduced in \cite{11,12}
are closely related to fracture functions 
in the same way that $F_2(x,Q^2)$ is related
to the usual parton densities. We wish to encourage an experimental
analysis in terms of such a direct parametrization and thus 
independent of the ``\pomeron{} exchange picture'' which, 
being non-perturbative,
is beyond present
theoretical control.

\begin{figure}[htb] \unitlength 1mm
\begin{center}
\dgpicture{138}{70}

\put(  3,20){\epsfig{figure=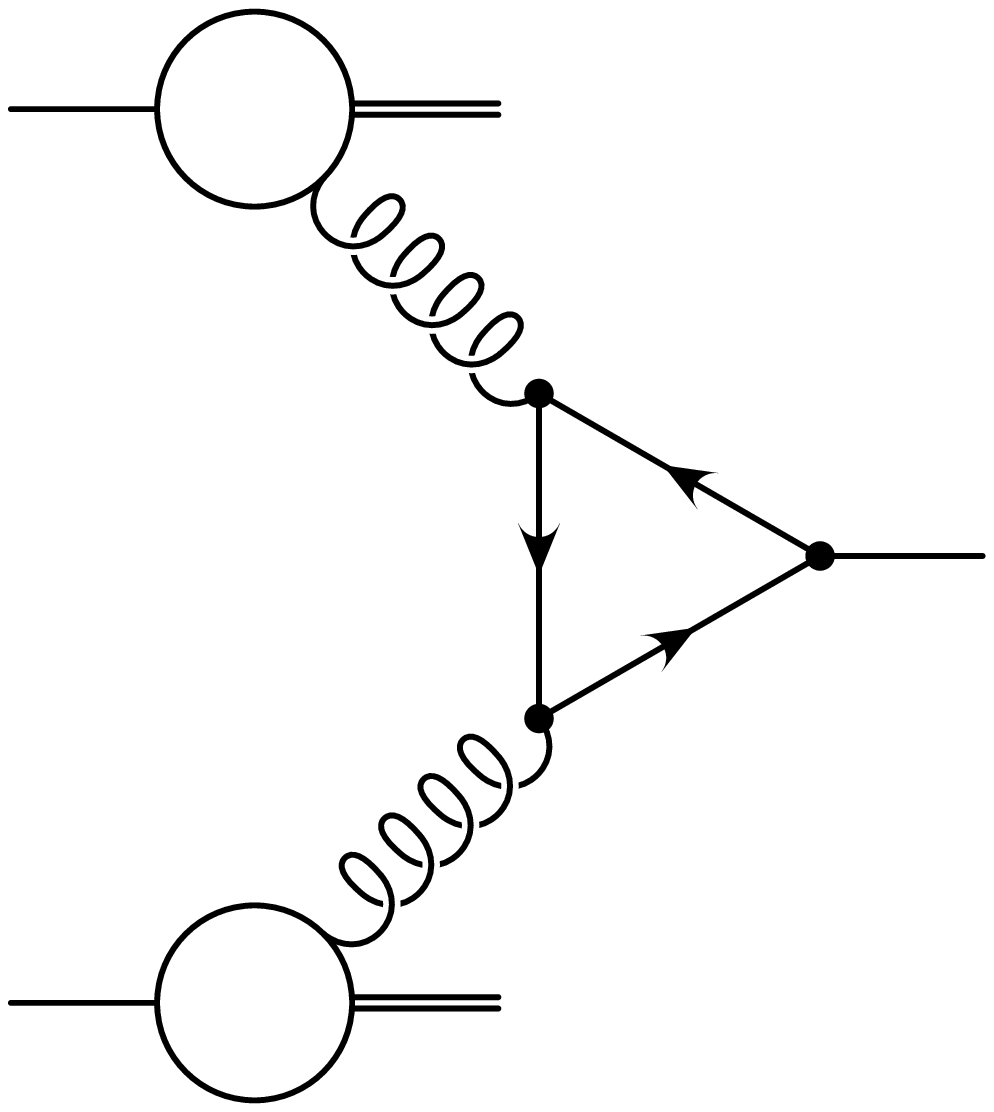,width=34mm}}
\put( 40,10){\epsfig{figure=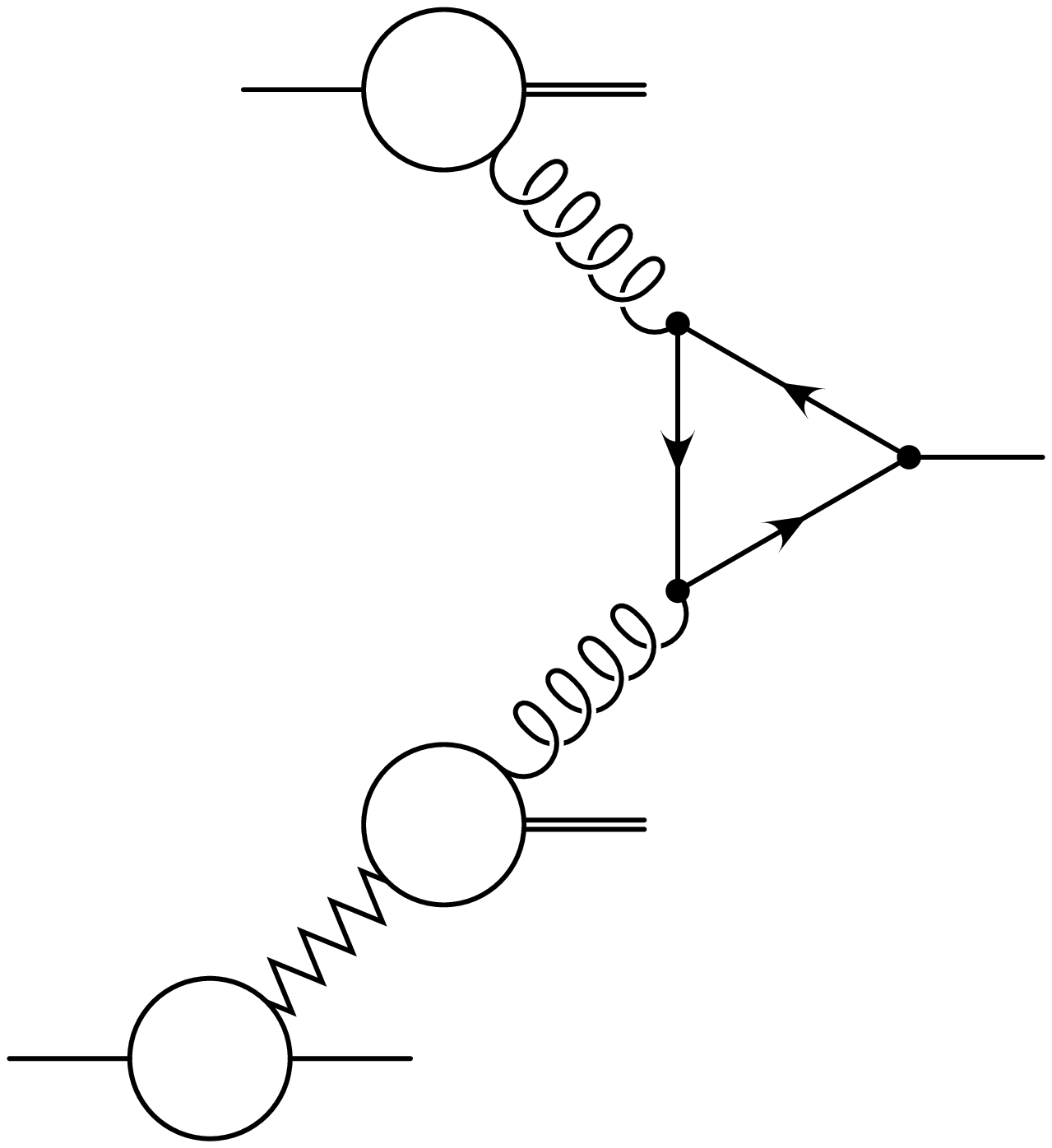,width=44mm}}
\put( 90,10){\epsfig{figure=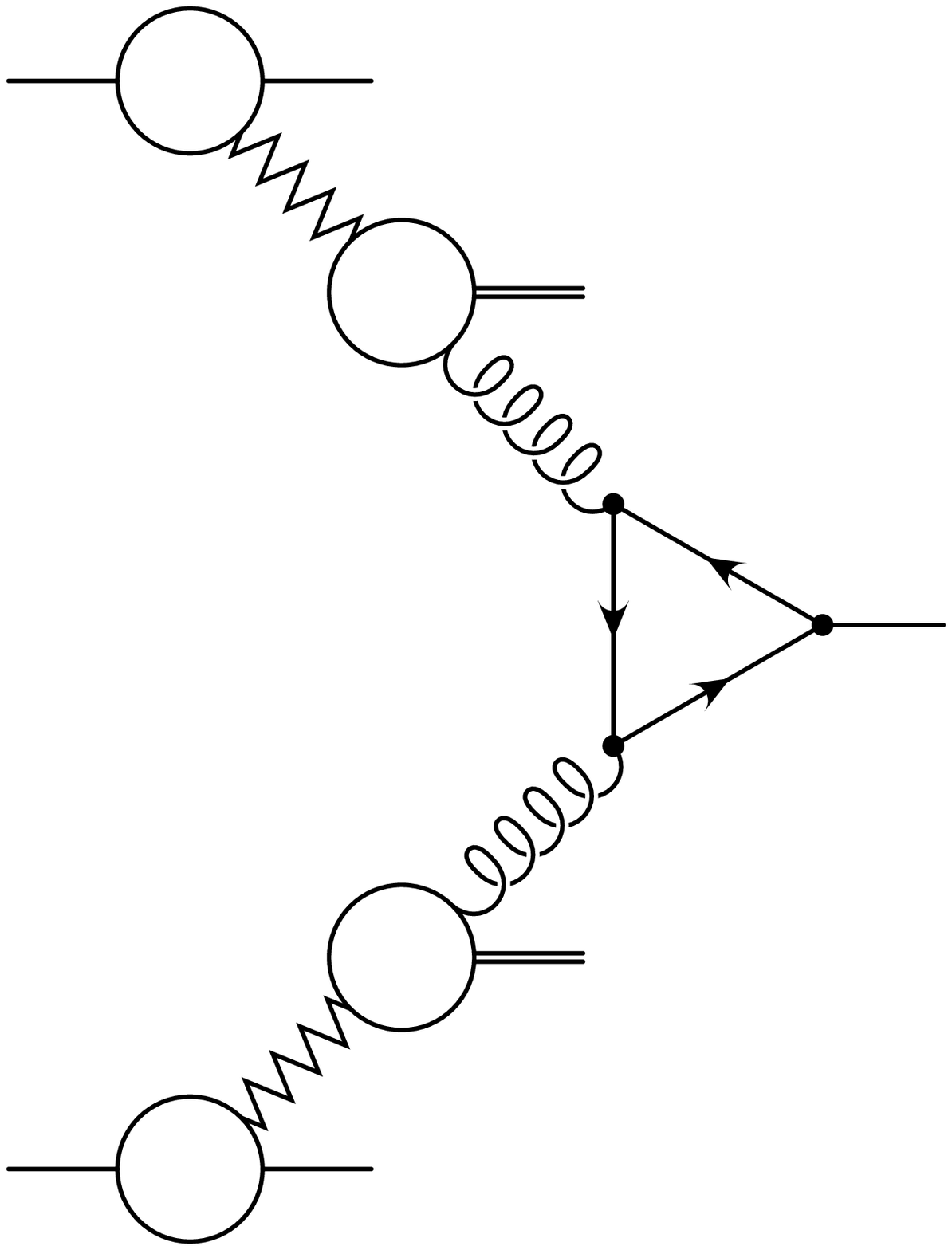,width=44mm}}

\put( 13,0){\it (a)}
\put( 50,0){\it (b)}
\put(100,0){\it (c)}

\end{picture}
\end{center}
\caption[]
{\label{pmech}{\it 
Generic diagrams for the total Higgs boson production cross
section (a), the single diffractive case (b) and the double diffractive
case (c). The quark in the triangle loop is the top quark. Single lines
stand for the incident proton and leading outgoing proton, and double lines
for the fragments of the proton and \pomeron.
}}   
\end{figure}

Figure~\ref{pmech}a shows the standard $gg$ fusion graph for Higgs production
via a top quark loop, giving the total 
cross section for $pp\rightarrow gg+X\rightarrow H+X$ in leading order, 
the process being diffractive or not.
Either one or both protons may stay essentially intact, 
giving rise to single (Fig.~\ref{pmech}b) and double (Fig.~\ref{pmech}c)
diffractive processes.
For a soft gluon content of the \pomeron, the double diffractive 
process has been 
studied in \cite{13}, and the formalism of Regge theory
has been applied in \cite{14}.
We base our study on the framework
for hard diffraction developed in \cite{15},
also employed in \cite{13}, and do not
make use of Regge theory, but instead assume the picture of the \pomeron{}
as an exchanged object carrying vacuum quantum numbers, 
with a parton content 
that can be measured and parametrized.

The flux $f_{{\Bbb P}/p}(x_{\Bbb P},t)$ of \pomeron{}s in the proton 
is given by the parametrization of Donnachie and Landshoff
\cite{16,17}. 
Actually, since the rapidity-gap criterion does not distinguish a leading 
proton from some other diffractively excited state, 
$f_{{\Bbb P}/p}$ should be increased relative to the
Donnachie--Landshoff value. Equivalently, one can follow
the practice of experimental groups to stick
to the flux of Refs.\ \cite{16,17} but allow the overall normalization
of the \pomeron{}'s parton densities to exceed the bounds imposed by a
naive momentum sum rule. For the parametrizations of the \pomeron's
gluon density\footnote{We do not include the model
of Buchm\"uller and Hebecker \cite{18} in our study, although 
it gives a first principles 
QCD formulation of diffractive events without employing
the concept of the \pomeron, because it is not clear how to define the
gluon density for diffractive events in this case. The concepts of
``rotation in colour space'' and ``colour-compensating soft gluon exchange'' 
of this model must probably 
be applied to 
the virtual top quark in the triangle loop. Because of {\em two} 
incident gluons, 
there should be an additional constraint related to the colour matching.} 
we use results from fits by the H1 \cite{6,7} and ZEUS
\cite{5,12} collaborations (parametrizations H1a, H1b, H1c and 
ZEUS\footnote{
The H1 parametrizations are based on fits of $F_2^{D(3)}$, 
whereas the ZEUS fits also take into account constraints from 
the photoproduction of jets. In the latter case, if only 
the information from $F_2^{D(3)}$ is used, 
it is possible to have a consistent set of parameters with a large
gluon content as well. To be definite, 
we use the central values given in \cite{5}.}) 
and moreover
employ recent results 
by Gehrmann and Stirling \cite{19} (parametrizations GS1 and GS2). 
In addition, to facilitate a comparison
with \cite{13}, we also include two parametrizations with a soft
gluon content (parametrizations S and SE, without and with scale evolution,
respectively). 
The parameters of the input distributions are shown in 
Tab.~\ref{pomtab}. In the cases where the parametrizations were not
available at arbitrary scales, the evolution has been carried out with the
standard leading-order Altarelli--Parisi equations with heavy flavour
thresholds at the single quark masses\footnote{The evolution 
for the fit of the parameters in the $F_2^{D(3)}$ analysis
of the H1 collaboration
has been done with $N_f=3$~flavours, and by adding charm via the photon--gluon
fusion process. In our case, the evolution span is sufficiently large 
in order to justify the treatment of the charm and bottom quarks
as massless flavours. The influence of changing $N_f$ from the
values used in our procedure to a fixed number of $3$
flavours is to increase the gluon density, because this decreases the
number of 
quark flavours into which the gluon can split. 
As a result, the cross section for
the single diffractive case increases by about $10$--$20\%$, 
and the one for the double diffractive case by about $20$--$40\%$.}
and $\Lambda_{\mbox{\scriptsize QCD}}=200\,$MeV for $4$~flavours.

\begin{table}
\caption[]%
{\label{pomtab}{\it{}Parametrizations of the \pomeron's parton content.
The input distributions are given by 
$\beta f_{g/{\Bbb P}}(\beta)=A\beta^B(1-\beta)^C$
for the gluon and 
$\beta f_{q/{\Bbb P}}(\beta)=D\beta^E(1-\beta)^F$ 
for $N_0$ light quark flavours 
$q$, $\overline{q}$ 
at the scale $\mu_0$. 
$\beta$ is the momentum fraction of the gluon in the \pomeron.
We have also included the
line types that are used in the plots later on.}}
\vspace{2mm}
\begin{center}
\begin{tabular}[htb]{|c||c|c|c|c|c|c|c|c|c|}
\cline{3-10}
\multicolumn{2}{c|}{}
&\rule[-2.5mm]{0mm}{8mm}$A$&$B$&$C$&$D$&$E$&$F$&
$N_0$&$\mu_0$\\ 
\hline
H1a & \dashline
       & $0$ & -- & -- & $0.189$ & $0.351$ & $0.355$ & 
       $3$ & $2\,\GeV$ \\ \hline
H1b & \dotline 
       & $9.80$ & $1$ & $1$ & $0.416$ & $1$ & $1$ & 
       $3$ & $2\,\GeV$ \\ \hline
H1c & \longdashline
       & $60.7$ & $7.99$ & $0.23$ & $0.260$ & $0.782$ & $1.21$ & 
       $3$ & $2\,\GeV$ \\ \hline
ZEUS & \dashdotline
       & $2.88$ & $1$ & $1$ & $0.48$ & $1$ & $1$ & 
       $2$ & $7.1\,\GeV$ \\ \hline
SE   & \dashdotdotdotline
     & $6$ & $0$ & $5$ & $0$ & -- & -- & 
       -- & $7.1\,\GeV$ \\ \hline
S    & \dashdashdotdotline
     & $6$ & $0$ & $5$ & $0$ & -- & -- & 
       \multicolumn{2}{c|}{no evolution} \\ \hline
GS1  & \dotdotline
     & \multicolumn{8}{c|}{cf.\ \cite{19}, model 1}\\ \hline
GS2  & \dashdotdotline
     & \multicolumn{8}{c|}{cf.\ \cite{19}, model 2}\\ \hline
\end{tabular}
\end{center}
\vspace{3mm}
\end{table}

\begin{figure}[htb] \unitlength 1mm
\begin{center}
\dgpicture{138}{100}

\put(  2, 5){\epsfig{figure=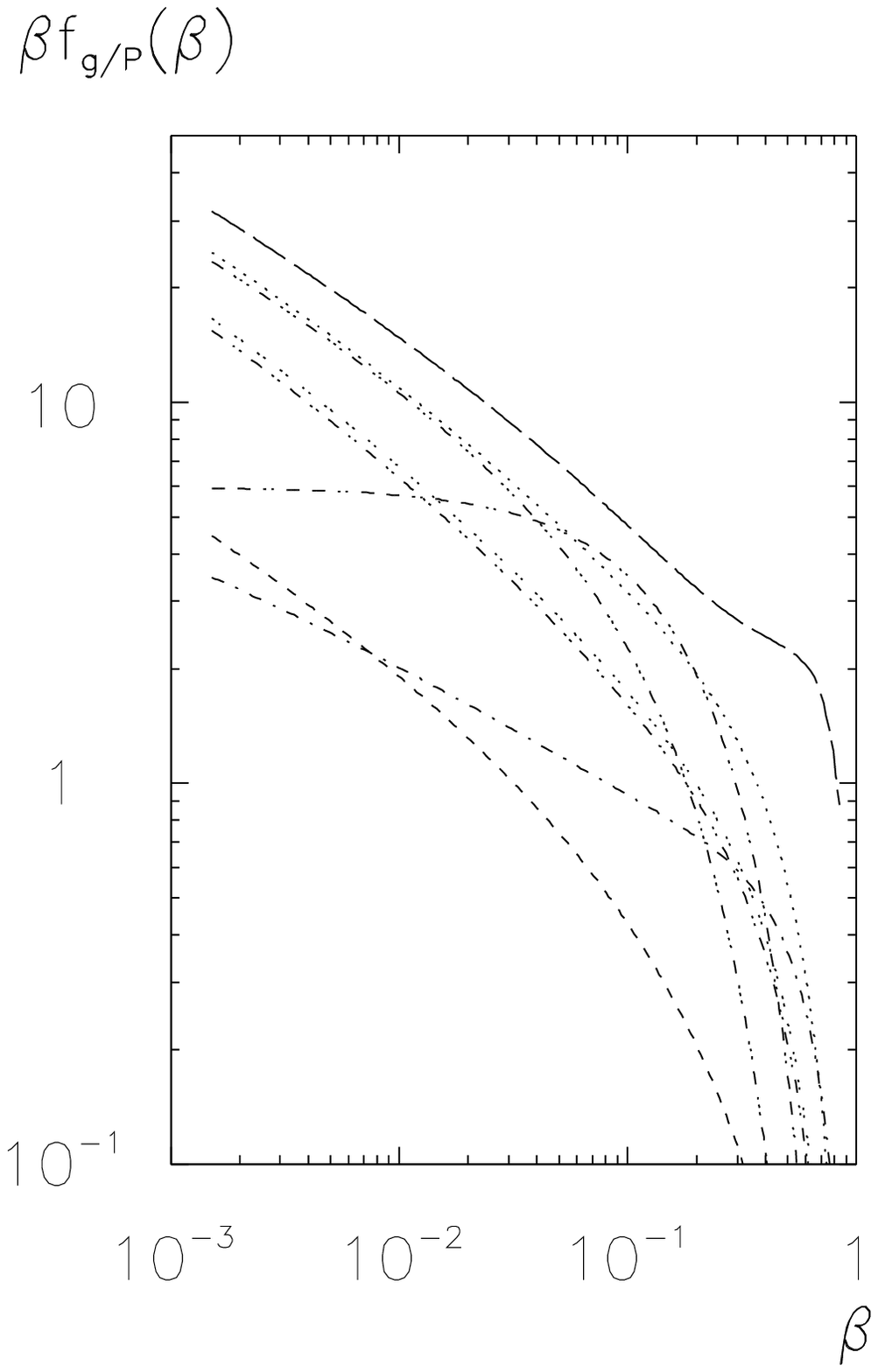,width=60mm}}
\put( 12, 0){\it (a)}

\put( 73, 5){\epsfig{figure=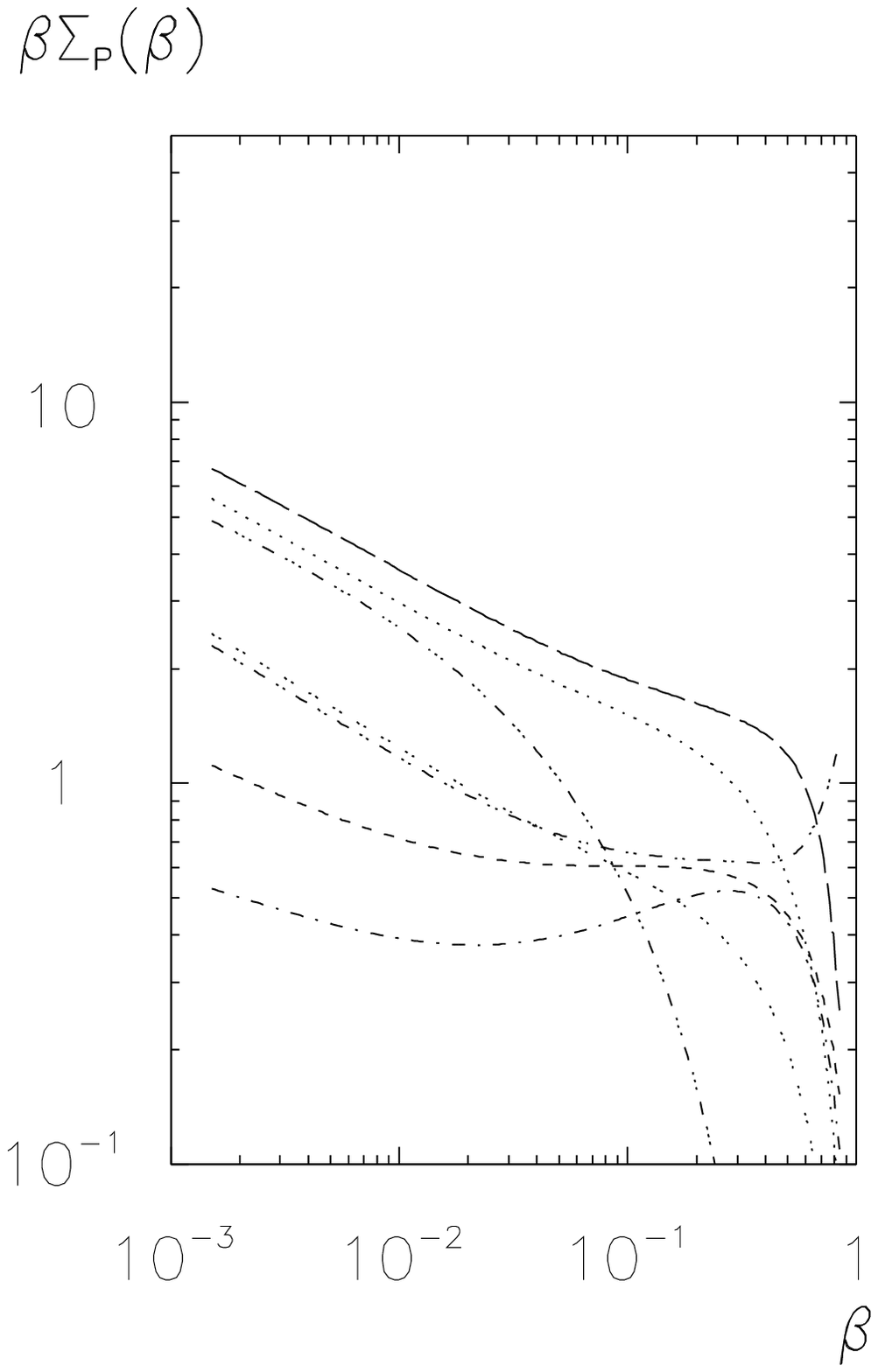,width=60mm}}
\put( 83, 0){\it (b)}

\end{picture}
\end{center}
\caption[]
{\label{gden}{\it
Parametrizations of the \pomeron's gluon density (a) and singlet quark 
distribution (b).
The line types are given in Tab.~\ref{pomtab}. 
}}   
\end{figure}

Figure~\ref{gden} shows the various gluon densities 
and the quark singlet distribution of the \pomeron{} 
at a scale of $100\,$\GeV.
Two of the H1 parametrizations (H1a and H1c) form a lower and upper bound
of all gluon parametrizations considered here; only the ZEUS parametrization
is smaller at small values of $\beta$. As we will see later on, the
\pomeron{} is probed mainly at large $\beta$, and
therefore the large spread of the parametrizations of about one order
of magnitude translates into a corresponding
variation in the calculated cross sections.
The quark singlet distribution is better constrained by the 
$F_2^{D(3)}$ measurement, and so the distributions have a smaller spread
at large $\beta$, where the present measurements are most sensitive.

The cross section for Higgs boson production via $gg$ fusion \cite{20} 
is given
by 
\begin{equation}
\label{xsect}
\sigma=\int_{\tau_H}^1\frac{\mbox{d}\xi}{\xi}\,g_1(\xi,\mu^2)\,
g_2\left(\frac{\tau_H}{\xi},\mu^2\right)\,\tau_H\,
\sigma_0\left(\frac{m_H^2}{m_{\mbox{\scriptsize top}}^2}\right).
\end{equation}
The quantity $\tau_H$ is defined by 
$m_H^2/E_{\mbox{\scriptsize CM}}^2$, where 
$E_{\mbox{\scriptsize CM}}$ is the
centre-of-mass energy of the collider. An explicit expression for
the parton-level cross section 
$\sigma_0(m_H^2/m_{\mbox{\scriptsize top}}^2)$ 
can be found, e.g., in \cite{21}.
For the total cross section and in the single diffractive case, 
both or one of the $g_i(\xi,\mu^2)$ 
are the gluon densities $f_{g/p}(\xi,\mu^2)$ of the proton, 
respectively,
which we choose to be the GRV leading-order parametrization~\cite{22}.
The other $g_i$'s, in the single and double diffractive case, 
are given by the convolution of the 
\pomeron{} flux factor $f_{{\Bbb P}/p}(x_{\Bbb P},t)$ and the 
parton densities of the \pomeron{} $f_{g/{\Bbb P}}(\beta,\mu^2)$:
\begin{equation}
\label{gflux}
g_i(\xi,\mu^2)=\int_{-\infty}^0 \mbox{d}t\,
\int_{\xi}^{\gamma(\xi,t)}
\frac{\mbox{d}x_{\Bbb P}}{x_{\Bbb P}}\,f_{{\Bbb P}/p}(x_{\Bbb P},t)\,
f_{g/{\Bbb P}}\left(\frac{\xi}{x_{\Bbb P}},\mu^2\right).
\end{equation}
The upper limit of the $x_{\Bbb P}$-integration is given by the condition
that the \pomeron's momentum fraction of the proton does not exceed
a certain fraction $x_{\Bbb P\mbox{\scriptsize max}}$, 
in order not to invalidate the parametrization used for
the \pomeron{} flux factor. Moreover, there is a kinematical 
limit depending on the momentum transfer $t=(p^\prime-p)^2$, where
$p$ and $p^\prime$ are the momenta of the incident and outgoing proton, 
respectively, so that
\begin{equation}
\gamma(\xi,t)=\max\left\{\xi,\,\min\left[x_{\Bbb P\mbox{\scriptsize max}},
\frac{-t}{2m_p^2}\left(\sqrt{1+\frac{4m_p^2}{-t}}-1\right)
\right]\right\},
\end{equation}
$m_p$ being the proton mass. In this expression we have neglected terms
of the order of $m_p/E_{\mbox{\scriptsize CM}}$ and
$\sqrt{-t}/E_{\mbox{\scriptsize CM}}$, reflecting the ambiguity in 
the massive case ($m_p\neq 0$, $t\neq 0$) of the
definition of the frame-dependent momentum fraction variable $x_{\Bbb P}$.

\begin{figure}[htb] \unitlength 1mm
\begin{center}
\dgpicture{138}{100}

\put(  2, 5){\epsfig{figure=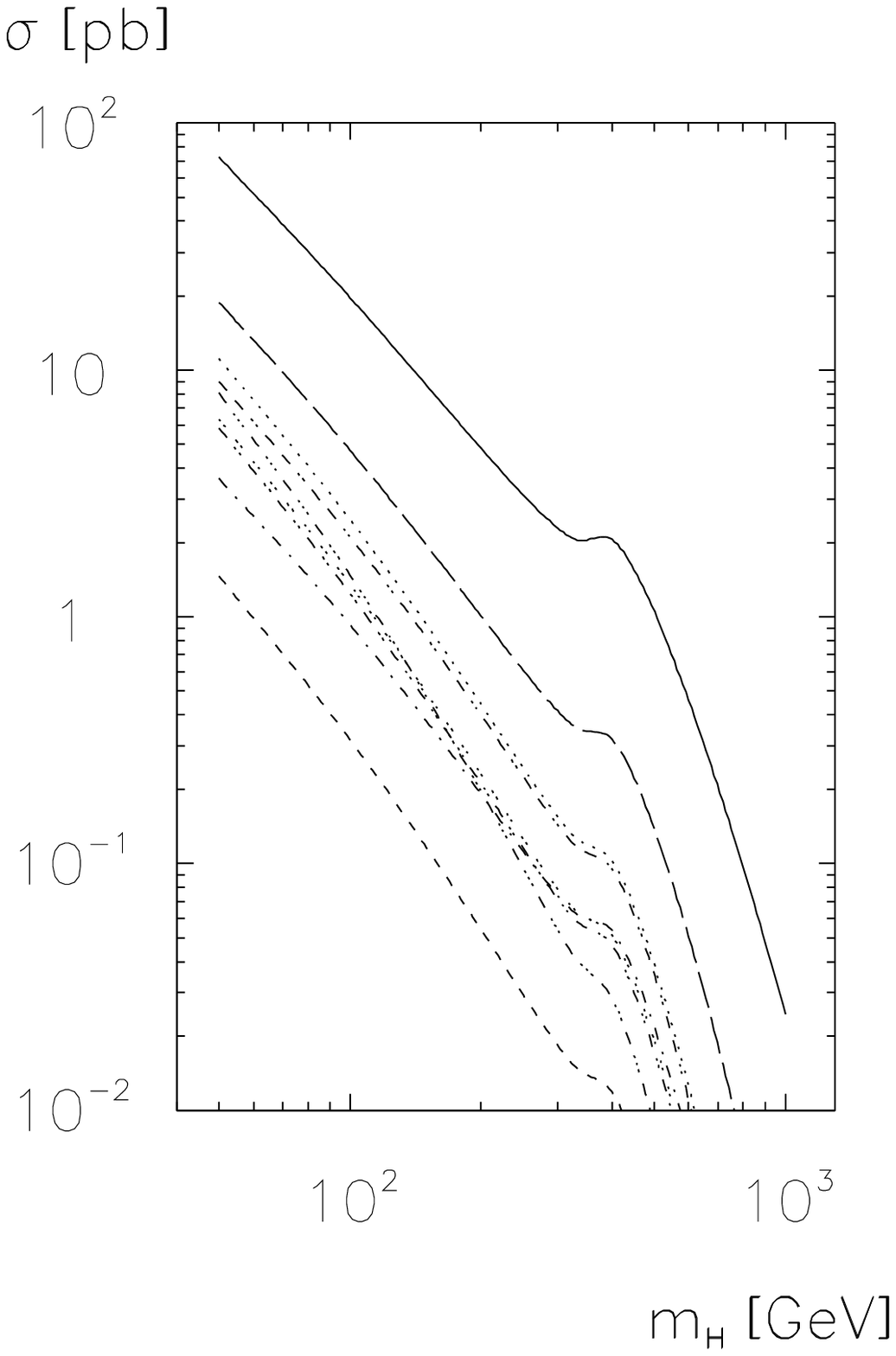,width=60mm}}
\put( 12, 0){\it (a)}

\put( 73, 5){\epsfig{figure=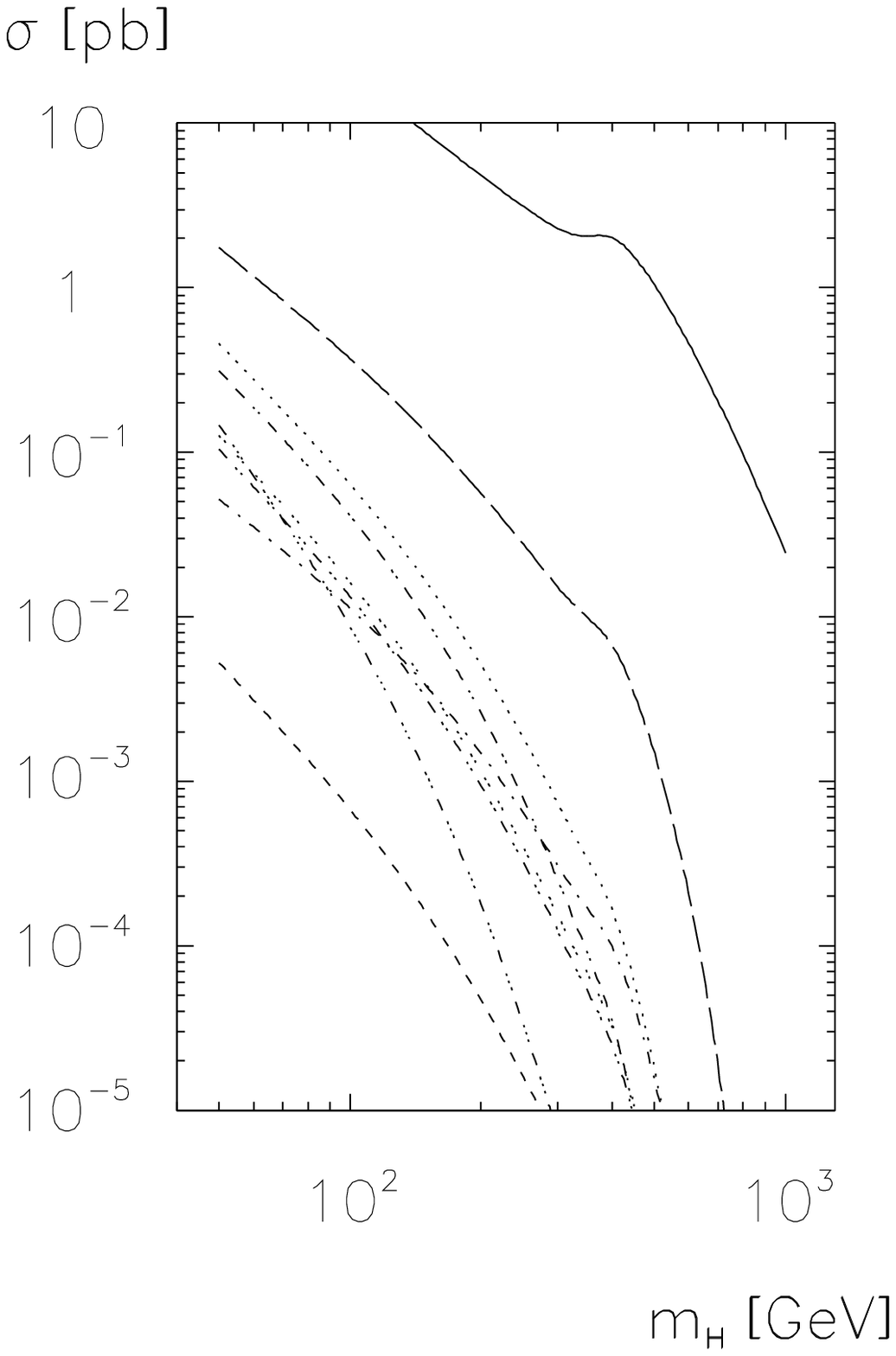,width=60mm}}
\put( 83, 0){\it (b)}

\end{picture}
\end{center}
\caption[]
{\label{mdep}{\it
Dependence of the single diffractive (a) and double diffractive (b)
Higgs boson production cross section on the Higgs boson mass.
The full line is the total cross section $pp\rightarrow gg+X\rightarrow H+X$. 
The other line types related to various gluon densities 
of the \pomeron{} are defined in Tab.~\ref{pomtab}. 
}}   
\end{figure}

For the numerical evaluation we have chosen 
$E_{\mbox{\scriptsize CM}}=10\,$TeV, $m_{\mbox{\scriptsize top}}=180\,$GeV
and
$x_{\Bbb P\mbox{\scriptsize max}}=0.1$.
Increasing the centre-of-mass energy to
$14\,$TeV results in an increase of all cross sections by about
a factor of~$2$.
The factorization scale $\mu$ and the renormalization scale
are set to the Higgs boson mass.
We use the leading-order running strong coupling
constant with $\Lambda_{\mbox{\scriptsize QCD}}=200\,$MeV for $4$~flavours.
The dependence of the cross section on the Higgs boson mass $m_H$ 
is shown in Fig.~\ref{mdep}a for the single diffractive 
case\footnote{We have subtracted the double diffractive cross section
from the single diffractive one, so what is actually plotted is the 
contribution where the proton which is modelled by the GRV gluon density
does not stay intact.} and in 
Fig.~\ref{mdep}b for the double diffractive case.
The shape of the mass dependence for the single diffractive case is similar 
to the non-diffractive case, although the cross section drops faster
for increasing Higgs boson mass.
In the mass range of $90\,\GeV<m_H<130\,\GeV$, the single diffractive cross
section for the H1c parametrization is about $25\%$ of the total cross section,
whereas most of the other parametrizations (excluding H1a) studied here give 
a fraction of the order of $10\%$.
Except for the H1c parametrization, the double diffractive cross section
is very small, less than about $0.1\,$pb.
Varying the gluon content of the ZEUS parametrization within the bounds
given in \cite{5} yields a variation of about $\pm 60\%$.
A comparison of the cross sections for the parametrizations S and SE shows
that the evolution of the parton densities gives,
in the double diffractive case, a large effect; 
however, this obviously depends on the input scale
chosen.
It should be noted that the diffractive cross section is not yet well
constrained by the measurements of the \pomeron{} structure function. 
For $m_H=100\,\GeV$, the spread is about a factor of $14$ in the single
diffractive case and about a factor of $600$ (!) in the double diffractive
case.
 
\begin{figure}[htb] \unitlength 1mm
\begin{center}
\dgpicture{138}{100}

\put(  2, 5){\epsfig{figure=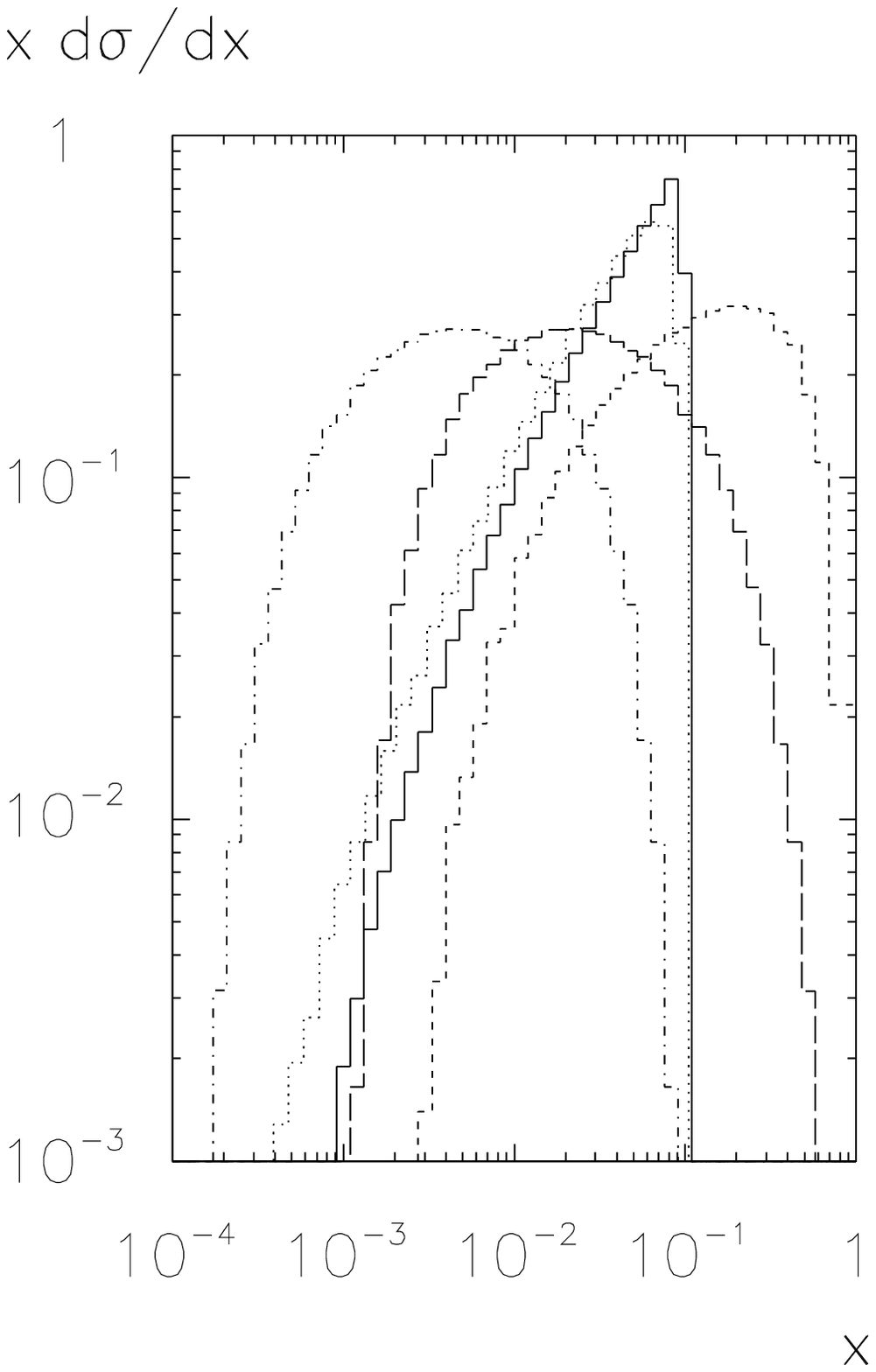,width=60mm}}
\put( 12, 0){\it (a)}

\put( 73, 5){\epsfig{figure=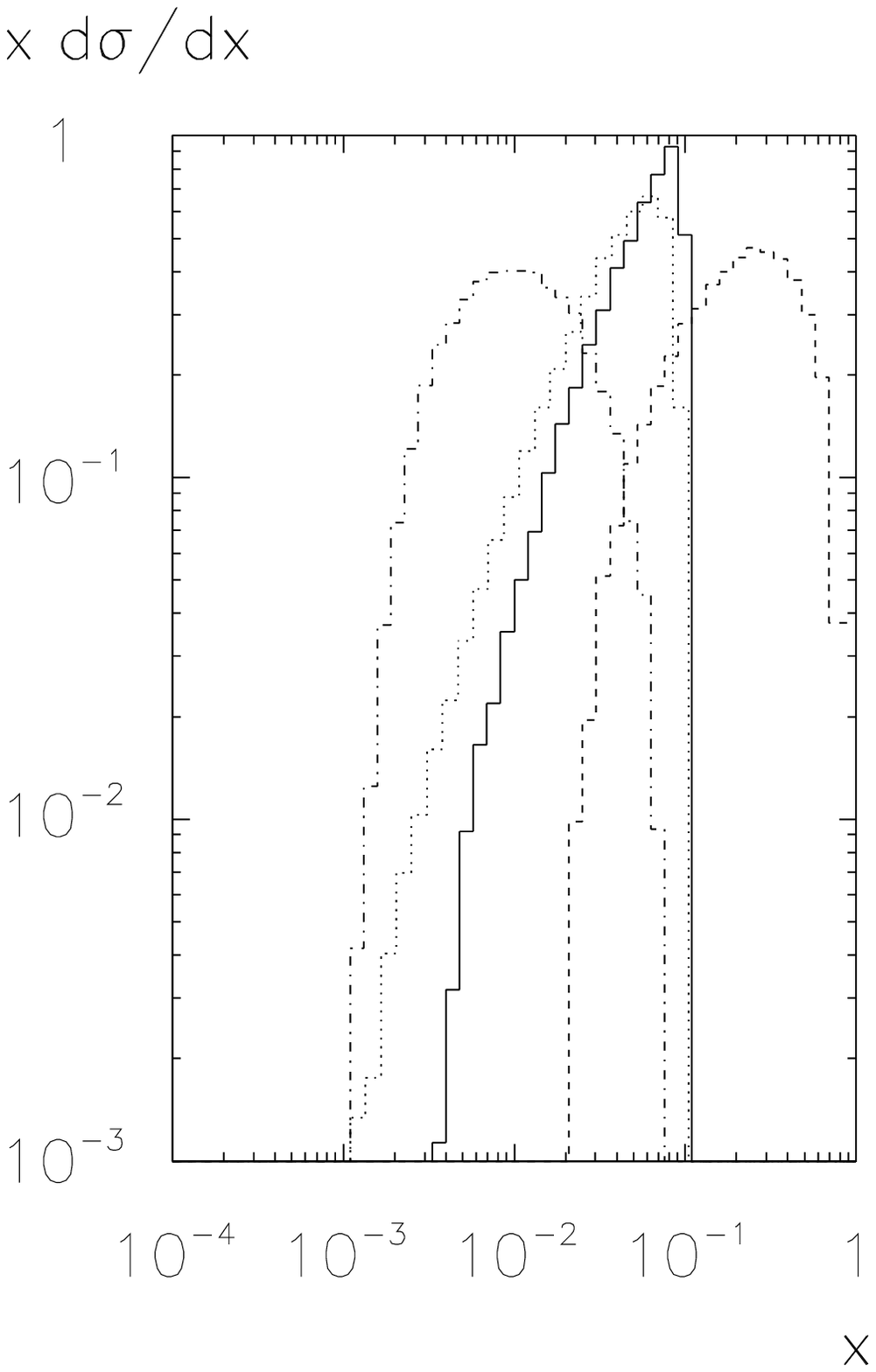,width=60mm}}
\put( 83, 0){\it (b)}

\end{picture}
\end{center}
\caption[]
{\label{mdist}{\it
Distributions $x\,\mbox{d}\sigma/\mbox{d}x$ for various variables $x$
in the single diffractive (a) and double diffractive (b) case
for a Higgs boson mass of $100\,\GeV$ and the H1b parametrization
of the gluon density. The integrated cross section $\sigma$ is normalized 
to~$1$. The variable $x$ stands for 
$x_{\Bbb P}$ \mbox{\rm[\fullline]},
$\beta$ \mbox{\rm[\dashline]},
$x_{\Bbb P}(1-\beta)$ \mbox{\rm[\dotline]},
$\xi_{ND}$ \mbox{\rm[\longdashline]},
$\xi_{D}$ \mbox{\rm[\dashdotline]}.
The ZEUS parametrization gives similar results.
}}   
\end{figure}

Differential distributions for the momentum fraction variables
$x_{\Bbb P}$, $\beta$, $\xi_{{\Bbb P}r}=x_{\Bbb P}(1-\beta)$, 
$\xi_{ND}$ and $\xi_{D}=x_{\Bbb P}\beta$
are shown in Fig~\ref{mdist}.
Here $\xi_{{\Bbb P}r}$ is the momentum fraction relative to the 
proton momentum carried by the remnant of the \pomeron, $\xi_{ND}$
is the momentum fraction of the gluon from the fragmenting proton
in the single diffractive case and $\xi_{D}$ is the momentum fraction of
the gluon from the \pomeron, relative to the momentum of the incoming proton.
The momentum fraction $\beta$ of the gluon in the \pomeron{} is
large, and the distribution $\beta\,\mbox{d}\sigma/\mbox{d}\beta$
peaks at about $0.2$.
The distribution of the momentum fraction $x_{\Bbb P}$ of the \pomeron{} in 
the proton,
being cut off by the explicit upper limit $x_{\Bbb P\mbox{\scriptsize max}}$,
has large contributions at large $x_{\Bbb P}$, although the \pomeron{}
flux behaves like $1/x_{\Bbb P}$.
The reason for this behaviour is that the invariant mass squared of the 
produced system in the hard scattering process is fixed
and quite large,
i.e.\ 
$\tau_H=m_H^2/E_{\mbox{\scriptsize CM}}^2=10^{-4}$ for $m_H=100\,\GeV$
and $E_{\mbox{\scriptsize CM}}=10\,$TeV. The momentum fraction $\xi$ 
to be supplied
by each of the protons is therefore of the order of $10^{-2}$ 
(cf.\ Fig.~\ref{mdist}). Since the gluon density of the \pomeron{} is
decreasing rapidly for increasing $\beta$, the \pomeron{} itself must carry
a comparably large fraction of the proton momentum.
This fact has the consequence that the momentum fraction $\xi_{{\Bbb P}r}$
carried by the ``\pomeron{} 
remnant'' relative to the proton momentum 
is, on the average, much larger than $10^{-2}$.
The rapidity gap which would be observed
in principle is between the beam direction and the \pomeron{} remnant. 
Assuming naively that the latter is simply a jet with longitudinal momentum
$\xi_{{\Bbb P}r}p$ and transverse momentum of the order of the transverse
momentum of the scattered proton, multiplied by $(1-\beta)$, then it turns
out that the distribution of the pseudorapidity $\eta$
of this jet peaks at values
of the order of $\eta=7$.
This means that hardly any gap will be observable in real 
experiments\footnote{It is to be expected that the same argument applies also to
the production of other systems with large invariant mass, such as 
jets and heavy quarks.}, 
even if fragmentation
effects are taken into account\footnote{The
\pomeron{} remnants will be dragged towards the hard process 
because of the missing colour
connection to the incident proton. However, 
this should not really influence the 
discussion, give or take one unit of rapidity.}.
Instead, for diffractive events, under the assumption that the 
scattered proton 
escapes detection, the signature will be a comparably small hadronic activity
in the forward direction, with the unobserved scattered proton carrying a large 
momentum fraction\footnote{If the mass of the Higgs boson is assumed
to be very small, of the order of $10\,\GeV$, then it turns out 
that the $x_{\Bbb P}$ and $\beta$ distributions are much broader, and that
the pseudorapidity distribution of the \pomeron{} remnant jet extends to smaller
values of~$\eta$.}.
In the double diffractive case, the measurement of the invariant mass of the
two-\pomeron{} system is in principle possible by measuring the energy of the
two scattered protons in forward spectrometers \cite{8}. If the \pomeron{}s
fragment, then
the measurement of the invariant mass
of the two-\pomeron{} system probably gives no constraint on the invariant
mass of the two-gluon system, because the gluons carry only 
a small fraction of the \pomeron{} momentum. 
There may, however, be a statistical correlation of $x_{\Bbb P}$ and $\xi_D$.
If the \pomeron{}s
do not fragment, then a measurement of the momenta of the scattered
proton gives the mass of the produced Higgs boson.
The cross section of this case, the so-called elastic part,
has been evaluated in 
\cite{14} in the framework of Regge theory and is found to be of the order
of $0.1\,$pb, which is of the same order of magnitude as what we get for the
total double diffractive case, including possible fragmentation 
of the \pomeron{}s. This situation should be clarified.

In order to make more reliable predictions for hard diffractive scattering
at $pp$ colliders, the gluon content 
of the \pomeron{} has to be determined much more precisely. The present 
experimental analyses cover a range from about $Q=3\,\GeV$ to 
$Q=10\,\GeV$, whereas the factorization scales to be employed
for $pp$ collider physics reach up to $1\,$TeV.
The gluon density at large scales is not very well constrained by the analysis
of $F_2^{D(3)}$
at small scales, because it contributes in \porder{\alpha_s} only,
and because of the large evolution span.
Additional constraints by studying heavy quark and
jet production at HERA and possibly at the Tevatron are needed to improve 
the situation.

The viability of the present study rests on the assumption that 
the simple factorization picture of eqs.~(\ref{xsect}) and
(\ref{gflux}) applies.
The analysis of a toy model shows that
the concept of factorization may break down for certain diffractive processes 
\cite{23}, and thus
the expression of the cross section as a convolution of parton densities
with a mass-factorized parton-level scattering cross section ceases
to be valid in principle, although it may still be
a good approximation in reality. 
This issue may be studied at present high-energy colliders.
The production of heavy quarks or jets at large $p_T$
in diffractive
events in $pp$ collisions is 
closely related to Higgs boson production studied in this paper,
in the sense that the produced system has a large invariant mass
and thus the momentum fractions probed in the proton and \pomeron{}
are large.
A comparison of the corresponding cross sections from the Tevatron
with
theoretical predictions would give a hint of what can be expected in 
the case of Higgs boson production at the LHC.

Finally, we wish to make some comments on the background to the process
studied here. As said before, 
the idea is to exploit the fact that the \pomeron{}
is a gluon-rich object, and thus background processes that are quark initiated
should be suppressed. 
We briefly consider two cases. One is the decay of the Higgs boson 
into $\tau^+\tau^-$, proposed in \cite{24}. It has been shown in 
\cite{25} that there is an overwhelming 
background from $t\overline{t}$ production that renders this channel 
unfeasible. Unfortunately, $t\overline{t}$ production proceeds
mainly via $gg$ fusion, and so this background is expected
to be large in the diffractive
case as well. It might be the case, 
however, that the $t\overline{t}$ production
cross section turns out to be smaller, because of the large invariant
mass of the $t\overline{t}$ pair. 
The other case we wish to comment on is the decay $H\rightarrow\gamma\gamma$.
As mentioned in the introduction, there are two types of backgrounds, 
the irreducible one from the production of photons via $q\overline{q}$,
$gg$ fusion and bremsstrahlung, and the reducible one, where jets fake 
photons. The latter background and the bremsstrahlung contribution 
can be reduced considerably by photon isolation cuts
\cite{26}. The direct production of
photon pairs in $q\overline{q}$ and $gg$ fusion has been studied in
\cite{27}, where it has been shown
that the $q\overline{q}$ process dominates.
This background might thus be reduced in 
the case of diffractive events.
It seems to be worth while to amend these  
qualitative remarks about the background by a thorough
quantitative study\footnote{We would like to remark that an effective gluon
density for hard diffractive processes may easily be implemented in existing
programs by parametrizing the convolution integral in eq.~(\ref{gflux}),
thus replacing a parton density by a fracture function.}. 

To summarize:
we have studied the production of the Standard Model Higgs boson at the 
LHC in single and double diffractive processes. 
Depending on the parametrization
of the gluon content of the \pomeron, the production cross section for
the single diffractive case may be sizeable and reach up to $25\%$
of the total Higgs boson production cross section, if the \pomeron{} is
gluon-dominated at large $\beta$ for small scales. 
However, the present parametrizations for
the gluon density of the \pomeron{} are not yet sufficiently precise
to allow for a reliable prediction.
Due to the large mass of the Higgs boson, the diffractive events
studied in this paper have only a ``small'' rapidity gap between the
leading proton and the 
\pomeron{} 
fragments, the latter having 
typically a very small transverse, but an extremely large 
longitudinal
momentum. Besides a leading proton
detected in a forward detector, the experimental signature
will be a comparably small
hadronic activity in the 
forward direction.

We wish to thank J.~Philips for discussions about
the parametrizations of the \pomeron's parton densities, 
L.~L\"{o}nnblad for clarifying comments on
rapidity gaps in electron--proton scattering, 
J.~Terron for a cross check of the evolution of the parton
densities,
L.~Trentadue for conversations about
fracture functions, and Z.~Kunszt for discussions about
the physics of the Higgs boson.

\newcommand{\scs}{\rm}

\newcommand{\bibitema}[1]{\bibitem[#1]{#1}}
\newcommand{\bibbeginlong}{\begin{thebibliography}{XXX\,1988\,\,}}
\newcommand{\bibbeginshort}{\begin{thebibliography}{XX}}

\newcommand{\bibitemb}[4]{\begin{sloppy}
{\bibitem[#1]{#1} {#2}, {#4.}}
\end{sloppy}}

\markh{References}
\bibbeginshort 

\addcontentsline{toc}{section}{References}

\bibitemb{1}
      {Z.~Kunszt and W.J.~Stirling}
      {}
      {in: Proceedings of the Large Hadron Collider ECFA Workshop,
       Aachen 1990, eds. G.~Jarlskog, D.~Rein}

\bibitemb{2}
      {ATLAS Collaboration}
      {}
      {Technical Proposal CERN/LHCC 94--43 (December 1994)}

\bibitemb{3}
      {CMS Collaboration}
      {}
      {Technical Proposal CERN/LHCC 94--38 (December 1994)}

\bibitemb{4} 
      {G.~Ingelman and P.E.~Schlein}
      {}
      {Phys.\ Lett.\ \underline{152B} (1985) 256}

\bibitemb{5} 
      {ZEUS Collaboration}
      {}
      {Phys.\ Lett.\ \underline{B356} (1995) 129}

\bibitemb{6} 
      {J.P.~Philips}
      {}
      {in: Proceedings of the Workshop 
       on Deep Inelastic Scattering, Paris 1995}

\bibitemb{7} 
      {H1 Collaboration}
      {}
      {DESY preprint in preparation}

\bibitemb{8} 
      {T.~Taylor, H.~Wenninger and A.~Zichichi}
      {}
      {preprint CERN-LAA/95-15 (June 1995)}

\bibitemb{9}
         {L.~Trentadue and G.~Veneziano}
         {Fracture Functions: an Improved Description of Inclusive 
          Hard Processes in QCD}
         {Phys.\ Lett.\ \underline{B323} (1994) 201}

\bibitemb{10}
      {D.~Graudenz}
      {}
      {Nucl.\ Phys.\ \underline{B432} (1994) 351}

\bibitemb{11} 
      {H1 Collaboration}
      {}
      {Phys.\ Lett.\ \underline{B348} (1995) 68}

\bibitemb{12} 
      {ZEUS Collaboration}
      {}
      {preprint DESY 95--93 (May 1995), 
       to be published in Z.~Phys.\ \underline{C}}

\bibitemb{13} 
      {O.~Nachtmann, A.~Sch\"afer and R.~Sch\"opf}
      {}
      {Phys.\ Lett.\ \underline{B249} (1990) 331}

\bibitemb{14} 
      {A.~Bialas and P.V.~Landshoff}
      {}
      {Phys.\ Lett.\ \underline{B256} (1991) 540}

\bibitemb{15} 
      {E.L.~Berger, J.C.~Collins, D.E.~Soper and G.~Sterman}
      {}
      {Nucl.\ Phys.\ \underline{B286} (1987) 704}

\bibitemb{16} 
      {A.~Donnachie and P.V.~Landshoff}
      {}
      {Phys.\ Lett.\ \underline{B191} (1987) 309}

\bibitemb{17} 
      {A.~Donnachie and P.V.~Landshoff}
      {}
      {Nucl.\ Phys.\ \underline{B303} (1988) 634}

\bibitemb{18} 
      {W.~Buchm\"uller and A.~Hebecker}
      {}
      {preprint DESY 95--77 (April 1995)}

\bibitemb{19} 
      {T.~Gehrmann and W.J.~Stirling}
      {}
      {Durham preprint DTP/95/26 (May 1995)}

\bibitemb{20} 
      {H.~Georgi, S.~Glashow, M.~Machacek and D.~Nanopoulos}
      {}
      {Phys.\ Rev.\ Lett.\ \underline{40} (1978) 692}

\bibitemb{21} 
      {A.~Djouadi, D.~Graudenz, M.~Spira and P.M.~Zerwas}
      {}
      {preprint CERN--TH/95--30 (February 1995)}

\bibitemb{22}
      {M.~Gl\"{u}ck, E.~Reya and A.~Vogt}
      {Dynamical Parton Distributions of the Proton
       and Small $x$ Physics}
      {Z.\ Phys.\ \underline{C67} (1995) 433}

\bibitemb{23}
      {A.~Berera and D.E.~Soper}
      {Diffractive Jet Production in a Simple Model with Applications to
       DESY HERA}
      {Phys.\ Rev.\ \underline{D50} (1994) 4328}

\bibitemb{24} 
      {R.K.~Ellis, I.~Hinchliffe, M.~Soldate and J.J.~van der Bij}
      {}
      {Nucl.\ Phys.\ \underline{B297} (1988) 221}

\bibitemb{25}
      {L.~DiLella}
      {}
      {in: Proceedings of the Large Hadron Collider ECFA Workshop,
       Aachen 1990, eds. G.~Jarlskog, D.~Rein}

\bibitemb{26} 
      {E.~Richter-Was, D.~Froidevaux, F.~Gianotti, L.~Poggioli, 
       D.~Cavalli and L.~Cozzi}
      {}
      {ATLAS Internal Note Phys-NO-048 (July 1995)}

\bibitemb{27}
      {P.~Aurenche, M.~Bonesini, L.~Camilleri, M.~Fontannaz and 
       M.~Werlen}
      {}
      {in: Proceedings of the Large Hadron Collider ECFA Workshop,
       Aachen 1990, eds. G.~Jarlskog, D.~Rein}

\end{thebibliography}

\end{document}